\documentclass[12pt,a4paper]{article}
\usepackage{graphicx}
\usepackage{times}
\title{\bf The IACOB spectroscopic database of Northern Galactic OB stars}
\author{S. Sim\'on-D\'iaz$^{1,2}$, N. Castro$^{1,2}$, M. Garcia$^{1,2}$, A. Herrero$^{1,2}$ and N. Markova$^3$\\
\vspace{1cm}\\
\normalsize $^1$ Instituto de Astrof\'isica de Canarias, E-38200 La Laguna, Tenerife, Spain.\\
\normalsize $^2$ Departamento de Astrof\'isica, Universidad de La Laguna, E-38205 La Laguna, Tenerife, Spain \\
\normalsize $^3$ Institute of Astronomy with  NAO, BAS, P.O. Box 136, 4700 Smolyan, Bulgaria}
\date{\mbox{}}
\begin{document}
\maketitle
\pagestyle{empty}
%
% WE REDEFINE THE plain LaTeX PAGESTYLE !!! 
% THIS PAGESTYLE WILL BE USED FOR THE FIRST PAGE ONLY !
%
\def\bull{\vrule height .9ex width .8ex depth -.1ex}
\makeatletter
\def\ps@plain{\let\@mkboth\gobbletwo
\def\@oddhead{}\def\@oddfoot{\hfil\tiny\bull\quad
``The multi-wavelength view of hot, massive stars''; 39$^{\rm th}$ Li\`ege Int.\ Astroph.\ Coll., 12-16 July 2010 \quad\bull}%
\def\@evenhead{}\let\@evenfoot\@oddfoot}
\makeatother
%
% AND DEFINE OUR MACROS FOR THE REFERENCE LIST
% I.E \beginrefer \refer and \endrefer
%
\def\beginrefer{\section*{References}%
\begin{quotation}\mbox{}\par}
\def\refer#1\par{{\setlength{\parindent}{-\leftmargin}\indent#1\par}}
\def\endrefer{\end{quotation}}
%
% BEGIN THE ABSTRACT CHAPTER WITH \noindent\small, ENCLOSE IT IN A GROUP
% AND BOLDFACE THE TITLE.
%
{\noindent\small{\bf Abstract:} 
We present the IACOB spectroscopic database, an homogeneous set
of high quality, high resolution spectra of Galactic O- and B-type
stars obtained with the FIES spectrograph attached to the Nordic
Optical Telescope. We also present some results from ongoing
projects using the IACOB database.
}
%
% NOW COMES THE MAIN BODY OF THE ARTICLE
%
\section{Introduction}

In an epoch in which we count on a new powerful generation of stellar atmopshere codes
that include most of the relevant physics for the modelling of massive OB stars, with (clusters of)
high efficiency computers allowing the computation of large grids of stellar models in more
than reasonable computational times, and with the possibility to obtain
good quality, medium resolution spectra of hundreds O and B-type stars in 
clusters outside the Milky way in just one snapshot (see e.g. the {\em FLAMES I \& II 
Surveys of Massive Stars}, Evans et al. 2008, 2010), the compilation of medium and high-resolution 
spectroscopic databases of OB stars in our Galaxy is becoming more and more important. 
With this idea in mind, two years ago we began to compile the IACOB spectroscopic 
database, aiming at constructing the largest database
of multiepoch, high resolution, high signal-to-noise ratio (S/N) spectra of Galactic 
Northern OB-type stars. The IACOB database perfectly complements the efforts also devoted in 
the last years by the GOSSS (P.I. Ma\'iz-Apellaniz; see also Sota et al., these proceedings)
and the OWN (P.I's Barb\'a \& Gamen, leading a multi-epoch, high-resolution spectroscopic survey
of Galactic O and WR stars in the Southern hemisphere; see Barb\'a et al. 2010) teams.  

\subsection{Characteristics of the IACOB database and present status}

We are using the FIES spectrograph\footnote{Detailed information about the NOT and FIES can be found
in http://www.not.iac.es} at the 2.56\,m Nordic Optical Telescope (NOT) in the Roque de 
los Muchachos observatory (La Palma, Spain) to compile spectra for the IACOB database. A summary of the 
instrumental configuration and observing dates (before Sept. 2010) is presented in Table~\ref{tab1}. 
Spectra of $\sim$100 stars with spectral types earlier than B2 and luminosity classes ranging 
from I (supergiants) to V (dwarfs) have been already compiled. The O-type targets were selected 
among those stars with V\,$\le$\,8 included in the GOS catalogue 
(GOSC, Ma\'iz-Apell\'aniz et al. 2004). The main part of the B-type stars sample correspond 
to the works presented in Sim\'on-D\'iaz (2010) and Sim\'on-D\'iaz et al. (2010). The final 
spectra usually have S/N~$\ge$~200. \\

% -----------------------------------------------------------------------
% --------------------           TABLE 2        ------------------------
% -----------------------------------------------------------------------
\begin{table}[]
\caption{\small General characteristics of the IACOB v1.0 spectroscopic database.
}\label{tab1}
\centering
{\small
\begin{tabular}{ll|cc}
\hline
\multicolumn{2}{c}{Instrumental configuration}  & \multicolumn{2}{c}{Observing run \& Dates} \\
\hline
Telesc.: NOT2.56\,m      & Spect. range:  3800\,-\,7000 \AA & 08\,A-D: 2008/11/05-08, & 10\,D: 2010/06/22 \\
Instr.: FIES           & Res. power:  46000                 & 09\,A-D: 2009/11/09-12, & 10\,E: 2010/07/15  \\
Mode: med-res              & Sampling:  0.03 \AA/pix            & 10\,A-C: 2010/06/05-07, & 10\,F: 2010/08/07 \\
\hline
\multicolumn{2}{c}{Spectral types: O4-B2 (I-V)} & \# stars: 105  & \# spectra: 720 \\
\hline
\end{tabular}
}
\end{table}
%-------------------------------------------------------------

\begin{figure}[h]
\centering
\includegraphics[width=17cm]{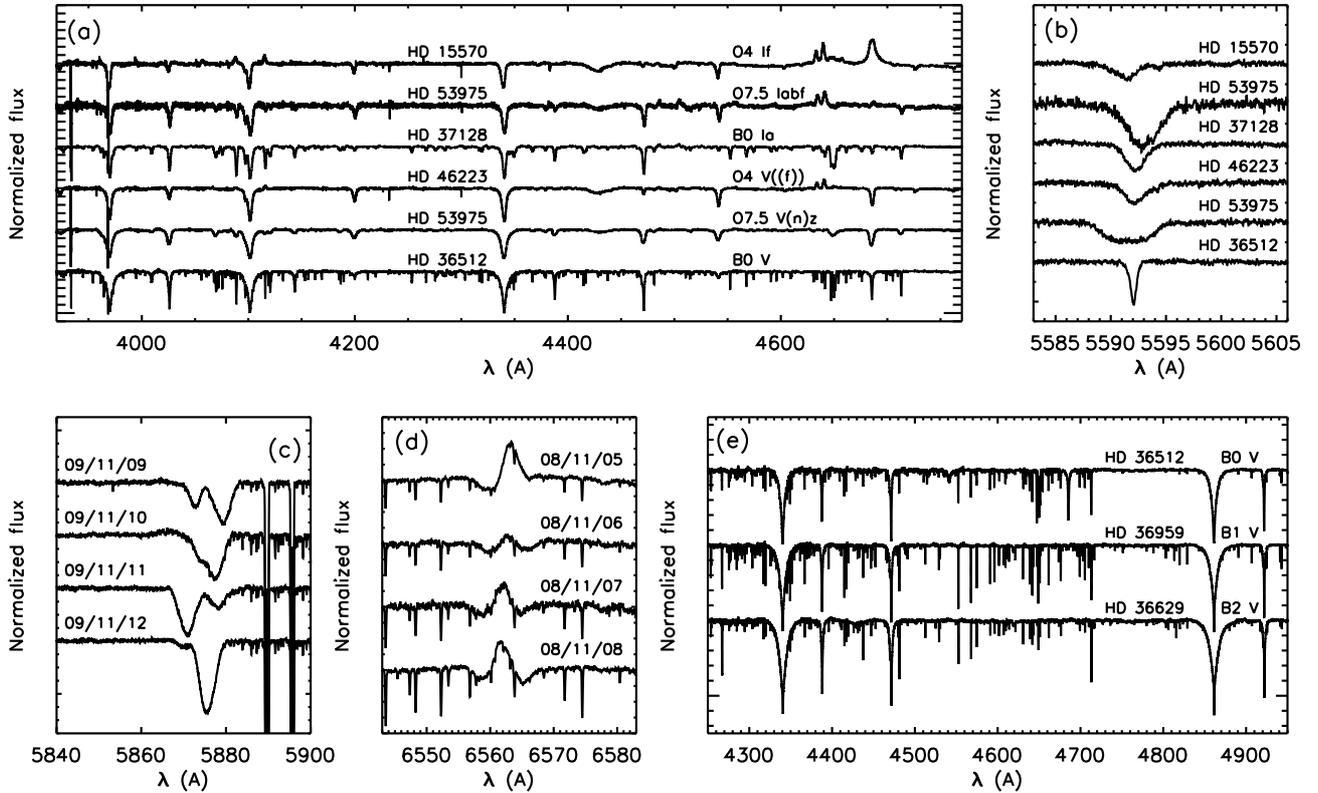}
\caption{{\small Illustrative examples of studies that can be performed using spectra from the 
IACOB database. (a) Spectra of a selected sample of MK standar stars centered in the spectral
region usually considered for the spectral classifications, 
(b) same stars in the region where the O\,{\sc iii}\,5592 line is located; this line
in the red part of the spectrum can be used to characterize the rotational and macroturbulent
broadening in O-type stars, where the Si\,{\sc iii}\,4552 line is no longer available. 
(c) HD\,1337, a double spectroscopic binary, observed in four consequetive nights, (d) HD\,37128, a B0\,Ia star in which has been detected
 strong variability in the H$\alpha$ line (this star is one of the selected targets for the investigation of
the macroturbulent broadening-pulsation connection in B supergiants, (e) Three narrow lined B-type 
stars from the Ori\,OB1 association, used in the reinvestigation of the chemical composition of 
young stars in the Orion star forming region.\label{fig_0}}}
\end{figure}

In Fig. \ref{fig_0} we present some examples of spectra in the IACOB database and studies that
can be performed with them.

\newpage

\section{Some ongoing projects using the IACOB spectroscopic database}

\subsection{Rotational velocities and macroturbulent broadening in OB stars}

\begin{figure}[h]
\centering
\includegraphics[height=17cm,angle=90]{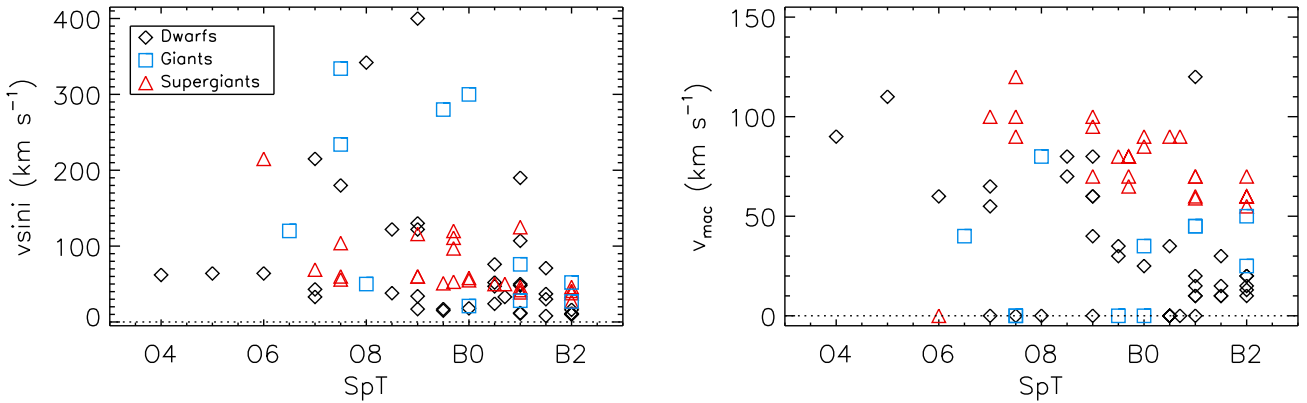}
\caption{{\small Rotational and macroturbulent velocities measured in the IACOB spectra. Double lined 
spectra were discarded. Not only the B Supergiants, but also the O-type stars show an important
macroturbulent broadening. The increasing trend of $v_{\rm mac}$ with SpT, previously found for 
B Supergiants (e.g. Fraser et al. 2010, and references therein), is continued in the O star domain. 
Note also the separation between dwarfs, giants and supergiants.\label{fig_1}}}
\end{figure}

We used the FT method (viz Gray 1976; see also Sim\'on-D\'iaz \& Herrero 2007, for a
recent application to OB stars) to disentangle the rotational and macroturbulent 
broadening and estimate their values in the whole sample of stars (see Fig. \ref{fig_1}). This analysis has
allowed us to confirm for the first time in a systematic way the presence of an
important non-rotational broadening in O-stars of all luminosity classes. If this effect is
not taken into account, it can significantly affect our v sini measurements. The origin of
this broadening is still not clear and presently under study (see below).

\subsection{Is macroturbulent broadening in OB stars related to pulsations?}

Part of the IACOB database consists of spectroscopic time series of 13 early-B Sgs (+ 2 early-B
dwarfs and 2 late-B Sgs). These observations were obtained to investigate the origin of the
macroturbulent broadening in OB stars and its possible relation to spectroscopic variability
phenomena and stellar pulsations.
First results show a tight correlation between the size of macroturbulent broadening 
and the line-profile variations present in all the early B-Sgs. More details about this 
investigation can be found in Sim\'on-D\'iaz et al. (2010), and in the talk contribution to these 
proceedings {\it ``Macroturbulent broadening in Massive Stars and its possible connection to
Stellar Oscillations''}.

\subsection{Increasing the statistics of OB stars with reliable determined stellar \& wind parameters}

One of the main aims driving the compilation of the IACOB spectroscopic database is to perform
a homogenous quantitative spectroscopic analysis of a statistically representative sample of
high-resolution, high quality spectra of stars with spectral types ranging from O4 to B2.
By means of these data we plan to refine the temperature calibration of Galactic O-type stars
and to address important questions such as, e.g. the weak wind problem, the mass discrepancy,
etc. The spectra will be analysed using the stellar atmopshere code FASTWIND (Puls et al. 2005), 
which allows us to create large grids of reliable spherical, NLTE, line blanketed models with 
winds for O and B stars of all luminosity classes in reasonable computational time-scales.
Two examples of results from preliminar analyses of the single lined objects from the IACOB
database are presented in Figs 3 and 4. Recently, we have established a collaboration with
the OWN group (P.I’s R. Barb\'a \& R. Gamen). They have been compiling for many years a similar
database with southern O stars, also investigating the multiplicty nature of apparently single objects.
We plan to join efforts to increase the sample of analyzed objects.

\begin{figure}[h]
{\scriptsize
\begin{minipage}{8cm}
\centering
\includegraphics[width=7.5cm]{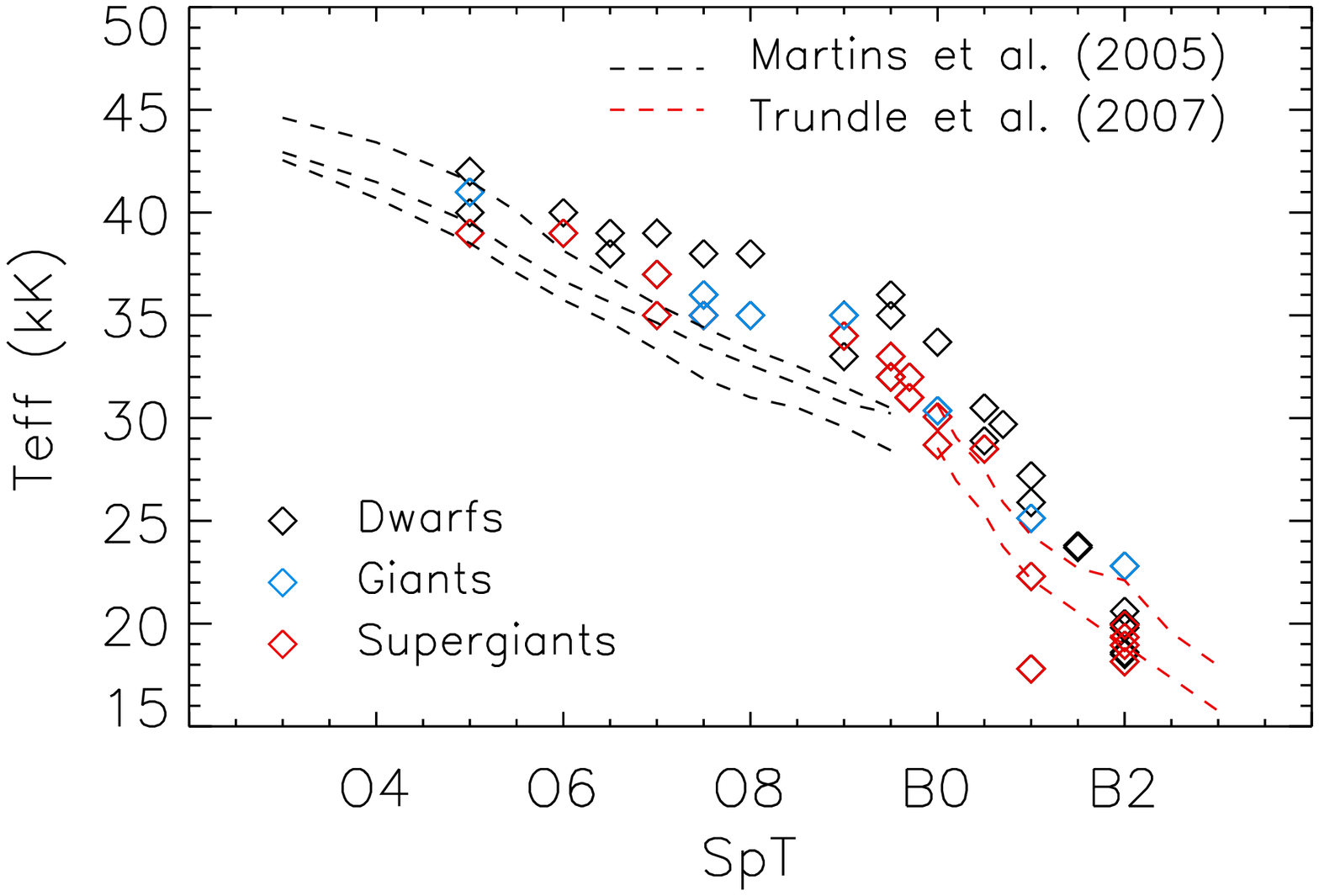}
\caption{\small{A homogeneous quantitative analysis of the IACOB spectra, accounting
for binarity/multiplicity, will help to better define the SpT\,-\,T$_{\rm eff}$ calibrations in the O to early-B star 
domain. Results from the preliminary analyses with FASTWIND, and comparison with two published
calibrations.} \label{fig_2}}
\end{minipage}
\hfill
\begin{minipage}{8cm}
\centering
\includegraphics[width=7.5cm]{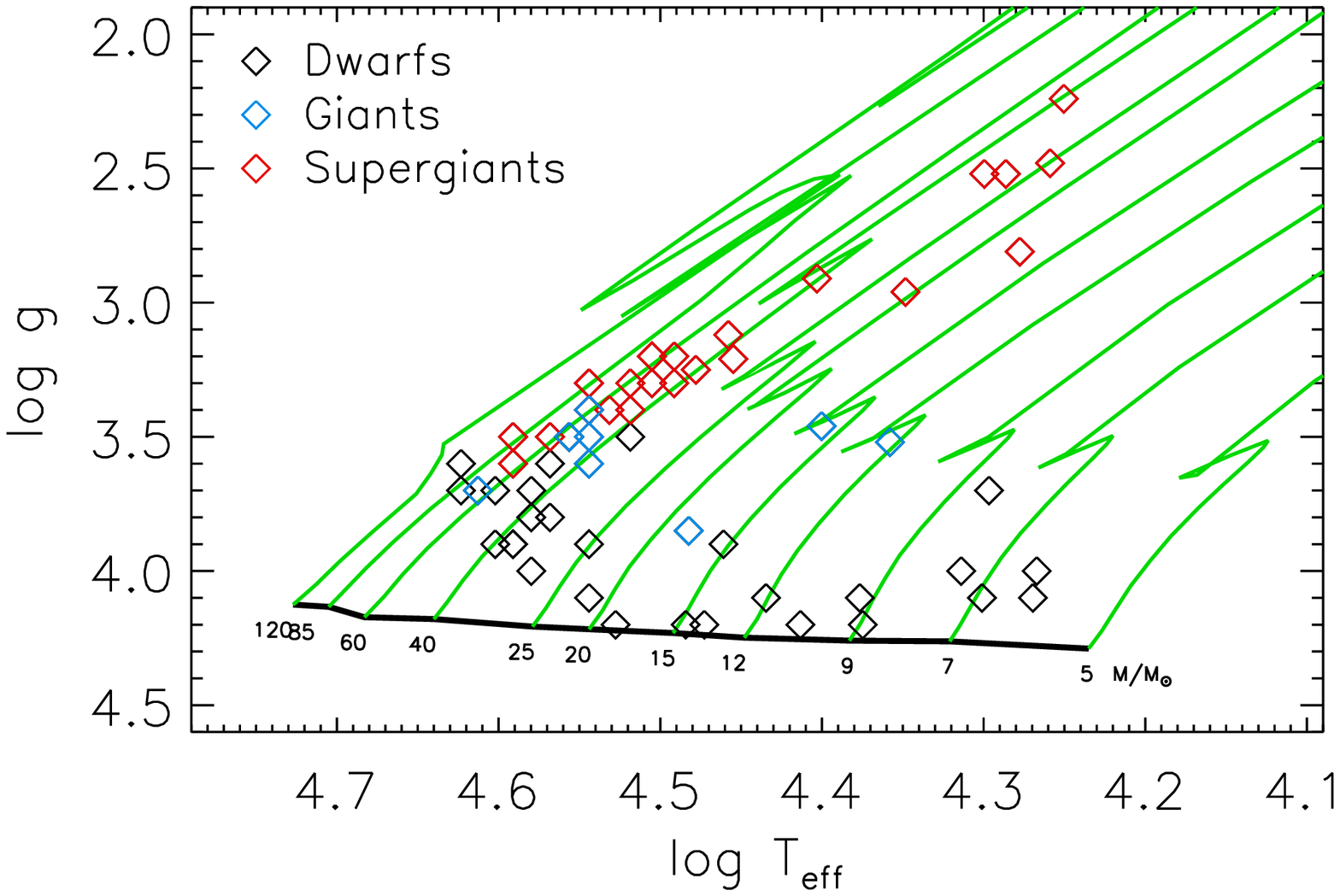}
\caption{{\small Position of the analyzed stars in the (log\,$g$, log\,T$_{\rm eff}$) 
diagram, along with the ZAMS (black, solid line) and evolutionary tracks from
Schaller et al. (1992) for stellar masses between 5 and 85 M$_\odot$ (green lines). 
Our analyses can help to better uncerstand the mass discrepancy problem.\label{fig_3}}}
\end{minipage}
}
\end{figure}

A similar analysis is in progress within the {\it FLAMES-II Survey of Massive Stars: ``The Tarantula survey''}
consortium (P.I. C. Evans). Around 1000 OB stars in 30\,Dor will be analyzed to
derive their rotational velocities, stellar and wind parameters, and N
abundances. The comparison of results from the two samples of stars (born in
environments with different metalicity) will be of great importance for our
undertanding of the physical properties and evolution of massive stars.

\subsection{Homogeneity of O and Si abundances in B-type stars in Ori OB1}

In Sim\'on-D\'iaz (2010), we used FASTWIND  to
perform a self-consistent spectroscopic analysis of a sample of 13 early B-type
stars from the Ori OB1 association observed with FIES@NOT. Main results
of this study, as part of a series of papers grouped under the title 
{\it ``The chemical composition of the Orion star forming region: stars, gas, and dust''}
can be found in the poster contribution to these proceedings with the same title.

\subsection{And more ...}

\begin{itemize}
\item A subsample of spectra from the IACOB (along with UVES and FEROS spectra from the ESO archive) have been recently
used within the {\it FLAMES-II Survey of Massive Stars: ``The Tarantula survey''} consortium to create a template 
atlas for spectral classification of medium resolution spectra of early type stars (Sana, Sim\'on-D\'iaz, Walborn
et al., in prep.)
\item Following ideas first presented in Markova et al. (2010), the IACOB database will be used (i) to investigate 
third parameter effects in the classification scheme developed by Walborn and co-authors to type O-stars and (ii) 
to recalibrate the logarithmic EW rations underlaying the Conti classification scheme.
\item We plan to use some of the IACOB spectra to analyze the feasibility of wavelet filtering to improve the
signal-to-noise ratio and, based on the sparsity of the spectra in the wavelet domain, to envisione the possibility to do
superresolution resting on the compressed sensing theory (Asensio Ramos \& L\'opez Ariste, 2010). 
\end{itemize}
 
\section{Future of the IACOB}

In the next semesters\footnote{The number of observed stars has already increased to $\sim$\,140 between
Sept. 2010 and the date of submission of this contribution}, we will continue with the 
compilation of spectra for the IACOB, observing stars with V$\le$8 in at least three epochs (more in the case of known
or newly detected binaries). Our idea is to make public the database via the Virtual 
Observatory in the next year. In the meantime, interested people can have access to 
the database under request to the author ({\tt ssimon@iac.es}). The complete list
of stars will be published in Sim\'on-D\'iaz et al., in prep.. We will acknowledge 
any observer who having obtained FIES spectra will like to add the spectra to the 
IACOB database after scientific exploitation.

%
% USE A SECTION WITHOUT NUMBER FOR THE ACKNOWLEDGEMENTS
%
\section*{Acknowledgements}
SSD, NC, MG and AH acknowledge finantial support by the Spanish MICINN (AYA2008-06166-C03-01 CSD2006-
00070). NM acknowledge financial support by the IAC and the Bulgarian NSF (DO02-85). We would like to also
thank the NOT people for their efficiency and kindness. SSD kindly thanks K. Uytterhoeven, J. Puls, C. Aerts, F.
Nieva, and N. Przybilla for their collaboration in some of the studies already published which made use
of spectra from the IACOB database. 
%
% BEGIN THE REFERENCE LIST WITH \beginrefer
% USE \refer BEFORE THE REFERENCES AND BEGIN A NEW PARAGRAPH AFTER THE 
% REFERENCE !
% DO NOT FORGET TO END THE LIST WITH \endrefer
%
\footnotesize
\beginrefer

\refer Aerts, C., Puls, J., Godart, M., \& Dupret, M.-A., 2009, A\&A, 508, 409

\refer Asensio Ramos \& L\'opez Ariste, 2010, A\&A, 509, 49

\refer Barb{\'a}, R.~H., Gamen, R., Arias, J.~I., et al.\ 2010, RMAyA, 38, 30 

\refer Evans, C.~J., Bastian, N., Beletsky, Y., et al., 2010, {\it IAU Symposium}, 266, 35 

\refer Evans, C.~J., Hunter, I.; Smartt, S., al., 2008, {\it The Messenger}, 131, 25 

\refer Fraser, M., Dufton, P. L., Hunter, I., \& Ryans, R. S. I., 2010, MNRAS, 404, 1306

\refer Gray, D. F. 1976, {\it The Observations and Analysis of Stellar Photospheres} (1st
ed.; New York: Wiley)

\refer Markova, N., Puls, J., Scuderi, S., et al., A\&A, submitted 

\refer Martins, F., Schaerer, D., \& Hillier, D. J., 2005, A\&A, 436, 1049

\refer Ma\'iz-Apell\'aniz, J., Walborn, N.~R., Galu\'e, H.~A.., \& Wei, L.~H., 2004, ApJS, 151, 103

\refer Puls, J., Urbaneja, M. A., Venero, R., et al., 2005, A\&A, 435, 669

\refer Schaller, G., Schaerer, D., Meynet, G., \& Maeder, A.\ 1992, A\&AS, 96, 269 

\refer Sim\'on-D\'iaz , 2010, A\&A, 510, 22

\refer Sim\'on-D\'iaz \& Herrero, 2007, A\&A, 468, 1063

\refer Sim\'on-D\'iaz, S., Herrero, A., Uytterhoeven, K., et al., 2010, ApJL, 720, 174

\refer Trundle, C., Dufton, P. L., Hunter, I., et al. 2007, A\&A, 471, 625

\endrefer           
\end{document}